\documentstyle[aps,prl,floats,graphicx,epsfig]{revtex}

\begin{document}
\draft
\twocolumn[\hsize\textwidth\columnwidth\hsize\csname @twocolumnfalse\endcsname

%
\title{Monopoles and fractional  vortices in chiral superconductors}
\author{G.E. Volovik  }

\address{  Low Temperature Laboratory, Helsinki
University of Technology, Box 2200, FIN-02015 HUT, Espoo, Finland\\
and\\ Landau Institute for Theoretical Physics, Moscow, Russia}

\date{\today} \maketitle
\
\centerline{Communicated by Olli V. Lounasmaa}
\begin{abstract}
We discuss two exotic objects which must be experimentally identified in
chiral superfluids and
superconductors. These are (i) the vortex with a fractional quantum number
($N=1/2$ in chiral superfluids,
and
$N=1/2$ and $N=1/4$ in chiral superconductors), which plays  the part of
the Alice string in
relativistic theories; and (ii) the hedgehog in the
$\hat{\bf l}$ field, which is the counterpart of the Dirac magnetic monopole.
These objects of different dimensions are topologically  connected. They
form the combined object which is called a nexus in relativistic theories. In
chiral
superconductors the nexus has magnetic charge emanating radially from the
hedgehog, while the
half-quantum vortices play the part of the Dirac string. Each of them supplies
the fractional magnetic flux to the
hedgehog, representing  1/4 of the "conventional" Dirac
string.  We discuss the
topological interaction of the superconductor's nexus with the `t
Hooft-Polyakov magnetic
monopole,  which can exist in Grand Unified Theories. The  monopole and the
hedgehog with the same magnetic charge are
topologically confined by a piece of the Abrikosov vortex. This makes the nexus
a natural trap for the magnetic monopole. Other properties of half-quantum
vortices and monopoles are discussed as well including fermion zero modes.

\end{abstract}
\
]

\section{Introduction}

Magnetic monopoles do not exist in classical electromagnetism.
Maxwell equations show
that the magnetic field is divergenceless,
$\nabla\cdot {\bf B}=0$, which implies that the magnetic flux
through any closed surface is zero: $\oint_S d{\bf S}\cdot{\bf B}=0$. If
one tries to construct
the monopole solution ${\bf B}= g {\bf r}/r^3$, the condition that
magntic field is nondivergent requires that magnetic flux $\Phi= 4\pi g$
from the monopole must
be acconpanied by an equal singular flux supplied to the monopole by an
attached
Dirac string. Quantum electrodynamics, however, can be successfully
modified to include
magnetic monopoles. In 1931 Dirac showed that the string emanating from a
magnetic monopole
becomes invisible for   electrons if the magnetic flux of the monopole is
quantized in terms of
the elementary magnetic flux
\cite{Dirac}:
\begin{equation}
4\pi g =n\Phi_0~,~\Phi_0={ hc \over  e}~,
\label{MagneticFluxQuantization}
\end{equation}
where $e$ is the charge of the electron.

In 1974 it was shown by `t Hooft \cite{Hooft} and Polyakov \cite{Polyakov},
that a magnetic
monopole with quantization of the magnetic charge according to
Eq.~(\ref{MagneticFluxQuantization}) can really occur as a physical object
if the $U(1)$ group of  electromagnetism is a part of the higher
symmetry group. The magnetic flux of a monopole in terms of the
elementary magnetic
flux coincides with the topological charge of the monopole: this is the
quantity
which remains constant
under any smooth deformation of the quantum fields. Such monopoles can
appear only in Grand
Unified Theories, where all interactions are united
by, say, the $SU(5)$ group.

\begin{figure}[!!!t]
\begin{center}
\leavevmode
\epsfig{file=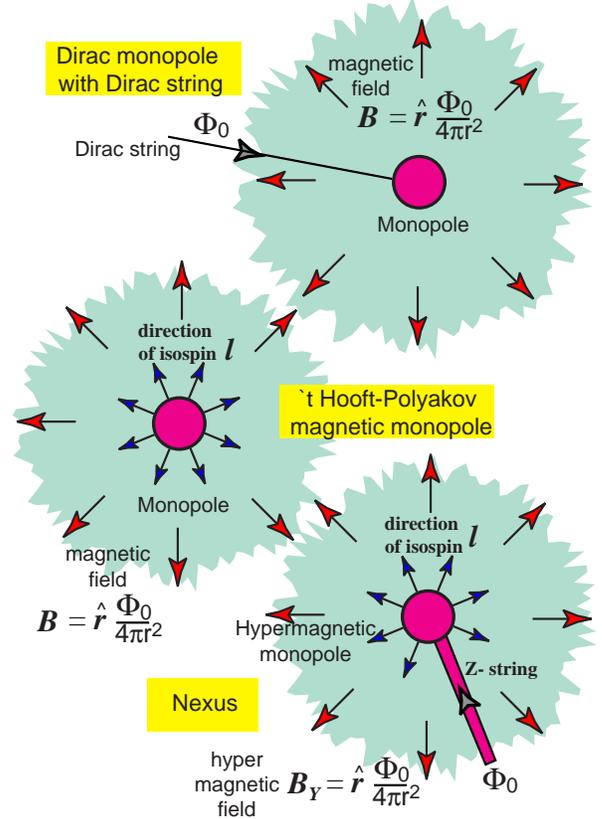,width=0.9\linewidth}
\caption[MonopoleNexus]
    {Dirac monopole, `t Hooft-Polyakov monopole, and electroweak monopole with
physical Dirac string.}
\label{MonopoleNexus}
\end{center}
\end{figure}

In the Standard Model of electroweak interactions such monopoles do not
exist, but the combined
objects monopole + string can be constructed without violating of the condition
$\nabla\cdot {\bf B}=0$. Further, following the terminology of
Ref.\cite{Cornwall} we shall call such combined object the nexus. In a
nexus the magnetic
monopole looks like a Dirac monopole but the Dirac string is  physical
and is represented by
the cosmic string. An example is the electroweak monopole discussed for the
Standard Model
(see Review \cite{AchucarroVachaspati}): the outgoing flux of the
hypermagnetic field is
compensated by the incoming hypercharge flux through the
$Z$-string (Fig.\ref{MonopoleNexus}.)

In condensed matter there are also topological objects,
which imitate magnetic monopoles. In chiral superconductors
their structure is
very similar to the nexus: it is the magnetic monopole combined either with
two  Abrikosov vortices, each carrying the flux $(1/2)\Phi_0$,   or with 4
half-quantum vortices,
each playing the part of 1/4 of the Dirac string. We also discuss the
interaction of such
topological defects in superconductors with the `t Hooft-Polyakov monopole.
If the latter
exists, then the nexus provides a natural topological trap for the
magnetic monopole.

\section{Symmetry groups}

The similarity between the objects in Standard Model and in chiral
superconductors stems from
the similar symmetry breaking scheme. In the Standard Model the local
electroweak symmetry
group $SU(2)_W\times U(1)_Y$ at high energy is broken at low energy to the
diagonal subgroup of the electromagnetism
$ U(1)_Q$, where $Q=Y-W_3$ is the electric charge. In amorphous chiral
superconductors
the relevant symmetry above the superconducting transition temperature
$T_c$ is $SO(3)_L\times U(1)_Q$,
where $SO(3)_L$ is
a global group of the orbital rotations.   Below  $T_c$ the symmetry is
broken to
the diagonal subgroup
$ U(1)_{Q-L_3}$. In high energy physics such symmetry breaking of the
global and local
groups to the diagonal global subgroup is called semilocal and the
corresponding topological
defects are called semilocal strings \cite{AchucarroVachaspati}. So in chiral
superconductors the strings are semilocal, while in chiral superfluids they are
global, since both groups in $SO(3)_L\times U(1)$ are global there.

If one first neglects the difference between the global
and local groups, the main difference between the symmetry breaking schemes
in high energy physics
and chiral superconductors is the discrete symmetry. It is the difference
between $SU(2)$ and
$SO(3)=SU(2)/Z_2$, and also one more discrete symmetry $Z_2$ which comes
from the coupling with
the spin degrees of freedom. This leads to the larger spectrum of the
strings and nexuses in
superconductors, as compared with the Standard Model.

\section{Fractional vortices in chiral superfluids/superconductors}

\subsection{Order parameter in chiral superfluids/superconductors}

The order parameter describing the  vacuum manifold in a chiral $p$-wave
superfluid ($^3$He-A) is the so called gap function, which in the
representation $S=1$ ($S$ is the spin momentum of Cooper pairs) and $L=1$
($L$ is the orbital angular momentum of Cooper pairs) depends linearly on spin
${\bf  \sigma}$ and momentum ${\bf k}$, viz.
\begin{equation}
\Delta({\bf k},{\bf r})=A_{\alpha i}({\bf r})  \sigma_\alpha k_i~,~
A_{\alpha i}= \Delta \hat d_\alpha( \hat e^{(1)}_i + i\hat e^{(2)}_i) ~.
\label{OrderParameter}
\end{equation}
Here $\hat{\bf d}$ is the unit vector of the spin-space anisotropy;  $\hat{\bf
e}^{(1)}$ and $\hat{\bf
e}^{(2)}$ are mutually orthogonal  unit vectors in the orbital space; they
determine the superfluid velocity of the chiral condensate ${\bf
v}_s={\hbar\over
2m}\hat e^{(1)}_i\nabla \hat e^{(2)}_i$, where $2m$ is the mass of the Cooper
pair; the orbital momentum vector is $\hat{\bf l}=\hat{\bf e}^{(1)}\times
\hat{\bf
e}^{(2)}$. The important discrete symmetry
comes from the identification of the points
$\hat{\bf d}$ ,  $\hat{\bf e}^{(1)}+i\hat{\bf e}^{(2)}$ and $-\hat{\bf d}$ ,
$-(\hat{\bf e}^{(1)}+i\hat{\bf
e}^{(2)})$: they correspond to the same value of the order parameter in
Eq.(\ref{OrderParameter}) and are thus physically indistinguishable.

The same order parameter desribes the chiral superconductor if the crystal
lattice influence
can be neglected, {\it e.g.} in an amorphous material. However, for
crystals the
symmetry group must
take into account the underlying crystal symmetry, and the classification
of the topological
defects becomes different. It is believed that   chiral
superconductivity occurs in
the tetragonal layered superconductor Sr$_2$RuO$_4$\cite{Rice}.  The
simplest representaion of
the order parameter, which reflects the underlying crystal structure, is
\begin{equation}
\Delta({\bf k},{\bf r})=({\bf d}\cdot{\bf \sigma})\left(\sin {\bf
k}\cdot{\bf a}({\bf r})+
i\sin {\bf k}\cdot{\bf b}({\bf r})\right)e^{i\theta}~,
\label{ChiralOP}
\end{equation}
 where $\theta$ is the phase of the order parameter; ${\bf a}$ and ${\bf
b}$ are the elementary
vectors of the crystal lattice. The order parameter is intrinsically
complex which cannot be
eliminated by a gauge transformation. This means that the time reversal
symmetry is
broken. Compare this with the structure of the
nonchiral $d$-wave superconductor in layered cuprate oxides, where the
order parameter is
complex only because of its phase:
\begin{equation}
\Delta({\bf k},{\bf r})=\left(\sin ^2{\bf k}\cdot{\bf a}({\bf r})-\sin^2{\bf
k}\cdot{\bf b}({\bf r})\right)e^{i\theta}~.
\label{DWaveOP}
\end{equation}
Because of the breaking of time reversal
symmetry in chiral crystalline superconductors, persistent electric current
arises
not only due to the phase coherence but also due to deformations of the
crystal:
\begin{equation}  {\bf j}=\rho_s\left( {\bf v}_s-{e\over mc}
{\bf A}\right) + Ka_i\nabla b_i~,~{\bf v}_s={\hbar \over 2m}\nabla \theta ~.
\label{SuperfluidCurrent}
\end{equation}
The parameter $K=0$ in $d$-wave superconductors.

The symmetry breaking scheme $SO(3)_S\times SO(3)_L\times U(1)_N\rightarrow
U(1)_{S_3}\times
U(1)_{N-L_3}\times Z_2$, realized by the order parameter in
Eq.(\ref{OrderParameter}), results in  linear topological defects (vortices or
strings) of group $Z_4$ \cite{VolMinTop}. Vortices are classified by the
circulation quantum number
$N=(2m/ h)\oint d{\bf r}\cdot {\bf v}_s$ around the vortex core.
Simplest realization  of the
$N$ vortex with integer $N$ is
$\hat{\bf e}^{(1)}+i\hat{\bf e}^{(2)}=(\hat{\bf x} +i\hat{\bf y})e^{iN\phi}$,
where $\phi$ is the azimuthal angle around the string. Vortices with even
$N$ are topologically
unstable and can be continuously transformed to a nonsingular
configuration.

\subsection{$N=1/2$ and $N=1/4$  vortices}

Vortices with a half-integer $N$ result from the above identification of
the points.
They are combinations of the $\pi$-vortex and
$\pi$-disclination in the $\hat{\bf d}$ field:
\begin{equation}
\hat{\bf d}=\hat{\bf x} \cos{\phi\over 2} +\hat{\bf y} \sin{\phi\over 2} ~,~
\hat{\bf e}^{(1)}+i\hat{\bf e}^{(2)}=e^{i\phi/2}(\hat{\bf x} +i\hat{\bf y}).
\label{HalfQuantumVortex}
\end{equation}

The $N=1/2$ vortex is the counterpart of Alice strings considered in
particle physics\cite{Schwarz}: a particle which moves around an Alice string
flips its charge. In $^3$He-A, the quasiparticle going around a $1/2$ vortex
flips its
$U(1)_{{\bf S}_3}$ charge, that is, its spin. This is because the ${\bf
d}$-vector,
which plays the role of the quantization axis for the spin of
a quasiparticle, rotates
by $\pi$ around the vortex, so that a quasiparticle adiabatically moving
around the vortex
insensibly finds its spin reversed with respect to the fixed environment.
As a consequence, several
phenomena ({\it eg.} global Aharonov-Bohm effect) discussed in the particle
physics literature have
corresponding discussions in condensed matter literature (see
\cite{Khazan1985,SalomaaVolovik1987} for
$^3$He-A and
\cite{March-Russel1992,Davis1994} in particle physics).

\begin{figure}
\begin{center}
\leavevmode
\epsfig{file=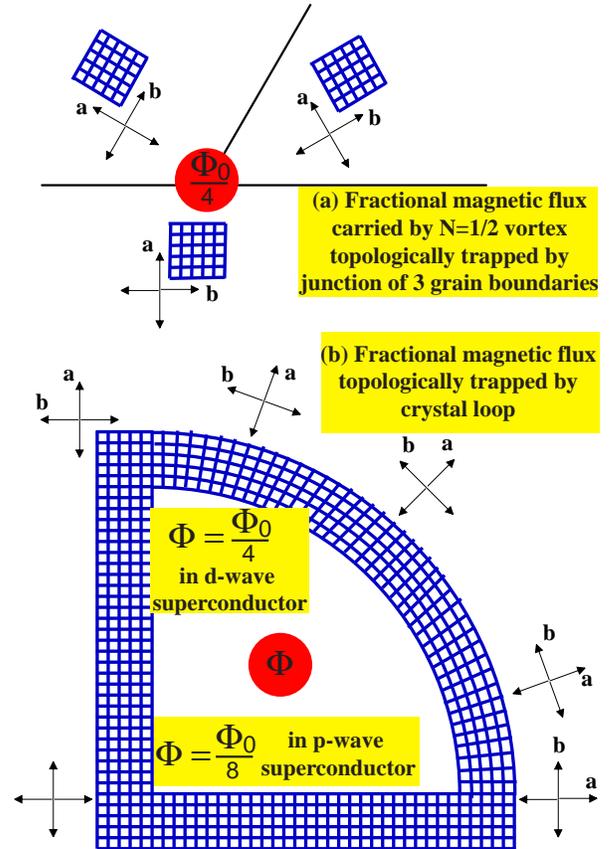,width=0.9\linewidth}
  \caption{(a) Experimentally  observed fractional flux, topologically
trapped by junctions of the
grain boundaries in high-$T_c$ superconductors. (b) Fractional flux
topologically trapped
by  loops of monocrystals with tetragonal symmetry.
The tetragonal crystal with $d$-wave pairing traps $N=1/2$ circulation
quanta and thus $1/4$ of
the quantum of magnetic flux. The same crystal with $p$-wave pairing traps
$N=1/4$ circulation
quanta and thus $1/8$ of the quantum of magnetic flux (if the parameter $K$ in
the deformation current in Eq.(\protect\ref{SuperfluidCurrent})
is neglected).
Note that in our notation the flux carried by a conventional $N=1$ Abrikosov
vortex in conventional superconductors is ${1\over 2}
\Phi_0$. The empty space inside the loop represents the common core of the
$\pi/2$ crystal
disclination and an $N=1/2$ or $N=1/4$ vortex.  }
  \label{1/2vortexDwave}
\end{center}
\end{figure}

\begin{figure}[!!!t]
\begin{center}
\leavevmode
\epsfig{file=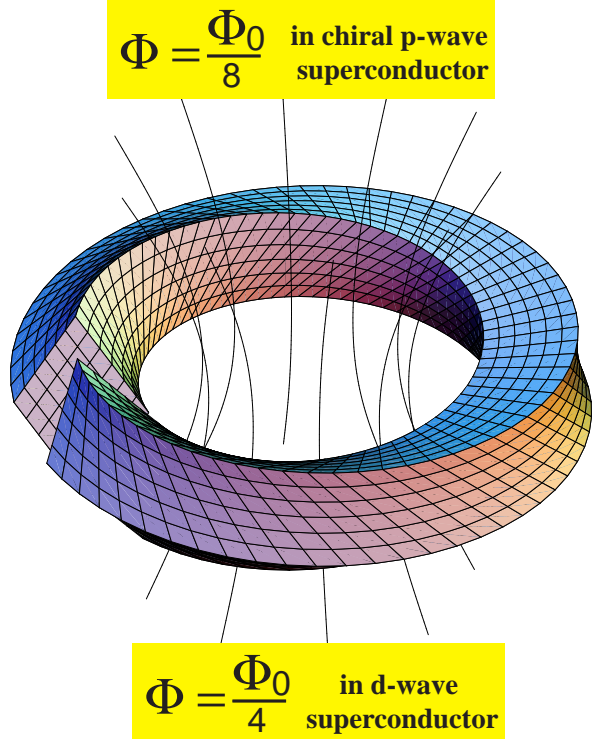,width=0.9\linewidth}
\caption[quarterVortex]
    {Fractional flux trapped by a loop of twisted wire with tetragonal cross
section. The wire is twisted by 90$^\circ$ and its surface behaves as
$1/2$ of the M\"obius strip: the surface transforms into itself only after 4
circulations around the loop. After 4
circulations the phase $\theta$ of the order parameter in
chiral
$p$-wave superconductor changes by $2\pi$. Thus the loop traps $1/4$ of the
circulation quantum. This corresponds to $\Phi_0/8$ of magnetic flux   if the
parameter
$K$ in the deformation current in Eq.(\protect\ref{SuperfluidCurrent}) is
neglected. }
\label{quarterVortex}
\end{center}
\end{figure}

In type II superconductors, vortices  with $N$ circulation quanta carry a
magnetic
flux $\Phi_N= (N/2)\Phi_0$; the extra factor $1/2$ comes from the Cooper
pairing nature of
superconductors. According to the London equations, screening of the
electric current far
from the vortex leads to the vector potential ${\bf A}= (mc/ e){\bf
v}_s$ and   to
the magnetic flux $\int d{\bf S}\cdot{\bf B}=\oint d{\bf r}\cdot{\bf
A}=(mc/ e)\oint d{\bf
r}\cdot{\bf v}_s = (N/2)\Phi_0$. Therefore, the conventional   $N=1$ Abrikosov
vortex in conventional
superconductors carries  ${1\over 2}
\Phi_0$, while the
$N=1/2$ vortex carries
$1/4$  of the elementary magnetic flux $\Phi_0$. The
vortex with  $N=1/2$ has been observed in high-temperature superconductors
\cite{Kirtley1996}: as
predicted in \cite{Geshkenbein1987} this vortex is attached to the tricrystal
line, which is the junction of
three grain boundaries (Fig.~\ref{1/2vortexDwave}.a).

Objects with
fractional
flux below $\Phi_0/2$ are also possible
\cite{VolovikGorkov1984}. They can arise if the time reversal symmetry is
broken \cite{Sigrist1989,Sigrist1995}.  Such fractional flux  can be
trapped by the crystal loop, which forms the topological defect, a
disclination: the
orientation of the crystal lattice continuously changes by $\pi/2$ around
the loop, Fig.~\ref{1/2vortexDwave}.b. The other
topologically similar loop can be constructed by twisting a thin wire by
an angle $\pi/2$ and then by
gluing the ends, Fig.~\ref{quarterVortex}.

Figs.~\ref{1/2vortexDwave}.b
and \ref{quarterVortex} illustrate fractional
vortices in the cases of
$d$-wave and chiral $p$-wave superconductivity in the tetragonal crystal.
Single-valuedness of the order parameter requires that the $\pi\over 2$
rotation of the
crystal axis around the loop must be compensated by a change of its phase
$\theta$. As a result the phase winding around the loop is $\pi$ for a
tetragonal
$d$-wave superconductor
and $\pi/2$ for a tetragonal $p$-wave superconductor. This means that the
loop of
$d$-wave
superconductor traps $N=1/2$ of the circulation quantum and thus
$(1/4)\Phi_0$ of the magnetic flux.

The loop of the chiral $p$-wave  superconductor traps 1/4 of the circulation
quantum. The magnetic flux trapped by the loop is obtained from the condition
that the electric current in Eq.(\protect\ref{SuperfluidCurrent}) is ${\bf
j}=0$ in superconductor. Therefore, the flux depends on the parameter
$K$ in the deformation current in Eq.(\protect\ref{SuperfluidCurrent}). In the
limit case when $K=0$ one obtains the fractional flux
$\Phi_0/8$. In the same manner $\Phi_0/12$ flux can be trapped if the
underlying crystal lattice
has hexagonal symmetry.

In $^3$He-B the experimentally identified nonaxisymmetric $N=1$
vortex \cite{Kondo1991} can be considered as a pair of $N=1/2$ vortices,
connected
by a wall \cite{Thuneberg1986,SalomaaVolovik1989,Volovik1990}.

\section{Nexus in chiral superfluids/superconductors}
\subsection{Nexus}

The  $Z$-string in the Standard Model, which has   $N=1$, is topologically
unstable, since
$N=0~(mod ~1)$. This means that the string may end at some point
(Fig.~\ref{MonopoleNexus}.c). The end
point, a hedgehog in the orientation of the weak isospin vector,
$\hat{\bf l}=\hat{\bf
r}$, looks like a Dirac monopole with
the hypermagnetic flux $\Phi_0$ in Eq.~(\ref{MagneticFluxQuantization}), if
the electric charge $e$
is substituted by the hypercharge \cite{AchucarroVachaspati}. The same
combined object of a
string and hedgehog,  the nexus,  appears in $^3$He-A  when the
topologically unstable
vortex with $N=2$ ends at the hedgehog in the orbital momentum field,
$\hat{\bf l}=\hat{\bf
r}$\cite{Blaha,VolMin}. In both cases the distribution of the vector
potential ${\bf A}$ of the
hypermagnetic field and of the superfluid velocity ${\bf v}_s$ field have
the same structure, if
one identifies ${\bf v}_s=(e/mc){\bf A}$. Assuming that the $Z$-string of
the Standard Model or
its counterpart in the electrically  neutral $^3$He-A, the
$N=2$ vortex,  occupy the lower half-axis
$z<0$, one has
\begin{eqnarray}
{\bf A} ={\hbar c\over 2e r}  {1-\cos\theta\over \sin\theta}\hat{\bf\phi},~
{\bf
B}=\nabla\times {\bf A}= {\bf B}_{mon} + {\bf B}_{string}  \\{\bf B}_{mon}=
{\hbar
c\over 2e}  {{\bf r}\over r^3}~,~ {\bf B}_{string}=-{hc\over e}\Theta(-z)
~ \delta_2({\bf \rho})~.
\label{DiracMonopoleField}
\end{eqnarray}
In amorphous chiral superconductors Eq.~({\ref{DiracMonopoleField})
describes the distribution
of the electromagnetic vector potential far from the nexus.
 The regular part of the magnetic field, radially propagating from the
hedghog, corresponds to a
monopole   with elementary  magnetic flux $\Phi_0= hc/e$, while the singular
part is concentrated in the core  of the  vortex, which supplies the
flux to the monopole \cite{MonopoleAPhase}.  This
is the doubly quantized
Abrikosov vortex, which is terminating on the hedgehog.

Because of the discrete symmetry group, the nexus structures in $^3$He-A  and
in amorphous chiral
superconductors are richer than in the Standard Model. The $N=2$ vortex can
split
into two
$N=1$ Abrikosov vortices or into
four $N=1/2$ vortices (Fig.~\ref{nexus}), or into their combination, provided
that the total topological charge
$N=0~(mod~2)$. So, in general, the superfluid velocity field in the
$^3$He-A nexus and the vector potential in its superconducting counterpart
obey
\begin{equation}
{\bf v}_s={e\over mc}   {\bf A}~,~  {\bf A}=  \sum_a  {\bf
A}^a~,
\label{SuperfluidVelocity}
\end{equation}
where ${\bf A}^a$ is the  vector potential of the electromagnetic field
produced
by the $a$-th string, {\it i.e.} the Abrikosov vortex  with the circulation
mumber $N_a$ terminating on the monopole, provided
that $\sum_a N_a=0~(mod~2)$.

\begin{figure}[!!!t]
\begin{center}
\leavevmode
\epsfig{file=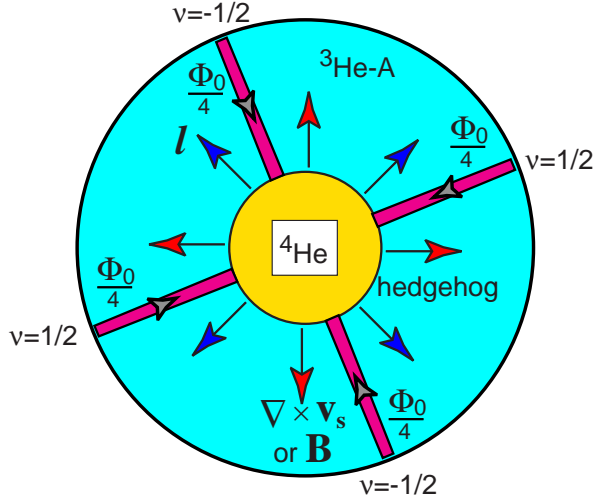,width=0.9\linewidth}
\caption[nexus]
    {Nexus in a small droplet of superfluid $^3$He-A: the hedgehog connecting
4 vortices with $N=1/2$ each. Blue arrows outward show the radial
distribution of
the orbital momentum
${\hat{\bf l}}$ field;  red   arrows outward illustrate  the radial
distribution
of superfluid vorticity
$\nabla\times {\bf v}_s$ or of magnetic field ${\bf
B}$ in the superconducting counterpart, a chiral superconductor. The magnetic
flux of the nexus $\Phi_0$ is supplied by 4 half-quantum vortices, each
carrying the flux $\Phi_0/4$ to the hedgehog. The charge
$\nu=\pm 1/2$ is the number of circulation quanta of the spin supercurrent
velocity
${\bf v}_{sp}$ around the half-quantum vortex. The stability of the monopole in
the center of the droplet is supported by the foreign body in the center, for
example by a cluster of
$^4$He liquid which provides the radial boundary condition for the ${\hat{\bf
l}}$-vector. }
\label{nexus}
\end{center}
\end{figure}

This is
similar to the other realization of the nexus
in  relativistic $SU(n)$
quantum field theories, for example in quantum chromodynamics, where
$n$ vortices of the group $Z_n$ meet at a center (nexus) provided the total
flux of
vortices adds to zero (mod $n$)
\cite{Cornwall,Chernodub,Reinhardt}.

\subsection{Nexus in a $^3$He droplet}

A nexus can be the ground state of  $^3$He-A in a droplet, if its radius
is less than 10 $\mu$m. In this case  the lowest energy of the nexus
occurs when all  vortices terminating on the monopole have
the lowest circulation number: this means that there must be four vortices with
$N_1=N_2=N_3=N_4=1/2$.

According to Eq.~(\ref{HalfQuantumVortex}) each half-quantum vortex is
accompanied by a spin disclination. Assuming that the $\hat{\bf d}$-field is
confined into a plane, the disclinations can be characterized by the winding
numbers
$\nu_a$ of the $\hat{\bf d}$ vector, which have values $\pm 1/2$ in
half-quantum
vortices. The corresponding spin-superfluid velocity
${\bf v}_{sp}$  is
\begin{equation}
{\bf v}_{sp}={2e\over mc}\sum_{a=1}^4 \nu_a {\bf A}^a~,~\sum_{a=1}^4 \nu_a=0~,
\label{SpinSuperfluidVelocity}
\end{equation}
where the last condition means the absence of the monopole in the spin
sector of
the order parameter. Thus we have $\nu_1=\nu_2=-\nu_3=-\nu_4= 1/2$.

If  $\hat{\bf
l}$ is fixed, the
energy of the
nexus in the spherical bubble of radius $R$ is determined by the kinetic energy
of mass and spin superflow:
\begin{eqnarray}
{1\over 2}\int dV \left(  \rho_s {\bf v}_s^2 +
\rho_{sp} {\bf v}_{sp}^2  \right)=\nonumber\\
{e^2\over 2m^2c^2} \int dV   (\rho_s+\rho_{sp})\left[({\bf
A}^1+{\bf A}^2)^2 +({\bf
A}^3+ {\bf A}^4)^2\right]   \nonumber\\
+{e^2\over m^2c^2} \int dV  (\rho_s-\rho_{sp}) ({\bf
A}^1+ {\bf A}^2) ({\bf
A}^3+ {\bf A}^4)~.
\label{Energy}
\end{eqnarray}

In the simplest case, which occurs in the ideal Fermi
gas, one has
$\rho_s
=\rho_{sp} $ \cite{VolWol}. In this case   the
$1/2$-vortices with positive spin-current circulation $\nu$ do not interact
with
$1/2$-vortices of negative $\nu$.  The energy minimum occurs when the
orientations of two positive-$\nu$ vortices are opposite, so that these two
${1\over 4}$ fractions of the Dirac strings form one line along the
diameter (see
Fig.~\ref{nexus}). The same happens for the other fractions with negative
$\nu$.
The mutual
orientations of the two diameters is arbitrary in this limit. However, in
real $^3$He-A, one has $\rho_{sp}<\rho_s $ \cite{VolWol}. If
$\rho_{sp} $ is slightly smaller than
$\rho_s$, the positive-$\nu$ and negative-$\nu$ strings repel each other,
so that
the equilibrium angle between them is $\pi/2$. In the extreme case
$\rho_{sp}\ll \rho_s $, the ends of four half-quantum vortices form
the vertices of a
regular tetrahedron.

To fix the position of the nexus in the center of the droplet, one must
introduce a spherical body inside, which will attach the nexus
because of the normal boundary conditions for the $\hat{\bf l}$ vector. The
body
can be a droplet of $^4$He immersed in the $^3$He liquid. For the mixed
$^4$He/$^3$He droplets,
obtained via the nozzle beam expansion of He gas, it is known that the
$^4$He component of the mixture does form a cluster in the central region
of the
$^3$He droplet
\cite{Vilesov}.

In an amorphous $p$-wave superconductor, but with preserved layered structure,
such a nexus will be formed in a  spherical shell. In the
crystalline Sr$_2$RuO$_4$ superconductor the spin-orbit coupling between
the spin
vector $\hat{\bf d}$ and the crystal lattice seems to align the
$\hat{\bf d}$ vector  along $\hat{\bf l}$\cite{Sigrist}. In this case the half
quantum vortices are energetically unfavorable and, instead of 4 half-quantum
vortices, one would have 2 singly quantized vortices in the spherical shell.

Nexuses of this kind can be formed also in the so called ferromagnetic Bose
condensate in optical traps. Such a condensate is described by a vector or
spinor
chiral order parameter \cite{Ho}.

\begin{figure}
\begin{center}
\leavevmode
\epsfig{file=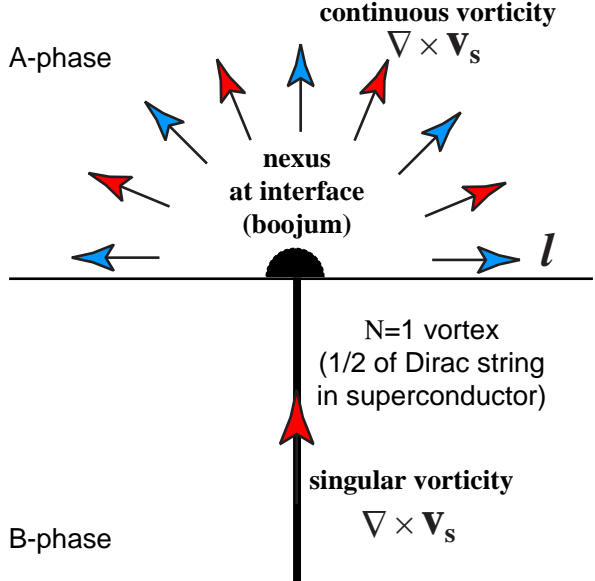,width=0.9\linewidth}
\caption[]{ $N=1$ vortex in $^3$He-B  terminating on a boojum -- the point
defect at the interface between $^3$He-B and $^3$He-A. Blue arrows show the
distribution of
$\hat{\bf l}$ on the
$^3$He-A side of the nexus, which resembles the gravimagnetic
monopole whose Dirac string is the B-phase vortex. In superconductors such
a nexus
accounts for 1/2  of the magnetic charge of the Dirac monopole, whose flux is
supported
by a single $N=1$ vortex on the B-phase side.}
\label{StringTermOnWall}
\end{center}
\end{figure}

\subsection{Nexus with fractional magnetic flux}

A nexus with fractional magnetic charge can be
constructed using geometry with several condensates.
Fig.~\ref{StringTermOnWall} shows the nexus pinned by the interface between
superfluid $^3$He-A and the nonchiral superfluid $^3$He-B. Due to the
tangential
boundary condition for the $\hat{\bf l}$ vector at the interface, the nexus
covers only half of the unit sphere. For the superconducting analogs  such
a nexus represents
the monopole with ${1\over 2}$ of the elementary flux $\Phi_0$. Thus on
the B-phase side, there is only one vortex with $N=1$ which terminates on
the nexus.

A monopole, which is topologically pinned by a surface or
interface is called a boojum \cite{Mermin}. The topological classification of
boojums is discussed in \cite{Misirpashaev,Volovik1992,Trebin}. In high energy
physics linear defects terminating on walls are called Dirichlet
defects
\cite{Dirichlet}.

\subsection{ Gravimagnetic monopole}

In addition to the symmetry breaking scheme there is another level of
analogies between superfluids/superconductors and quantum vacuum. They are
related
to the behavior of quasiparticles in both systems. In chiral superfluids
quasparticles behave as chiral fermions living in the effective gauge and
gravity fields, produced by the bosonic collective modes of the superflud
vacuum
(see Review\cite{PNAS}). In particular, the superfluid velocity acts on
quasiparticles in the same way as the  metric element $g^{0i}=-v_s^i$ acts  on
a relativistic particle in Einstein's theory. This element
${\bf g}=-g^{0i}$ plays the part of the vector potential  of the gravimagnetic
field
${\bf B}_g=\nabla\times {\bf g}$.

For the nexus in Fig.~\ref{StringTermOnWall}  the
${\hat{\bf l}}$ vector, the
superfluid velocity ${\bf v}_s$, and its "gravimagnetic field", i.e.  vorticity
${\bf B}_g$, on the A-phase side are
\begin{equation}
{\hat{\bf l}}={\hat{\bf r}}~,~{\bf v}_s= {\hbar\over 2m_3} {1-\cos\theta \over
r\sin\theta}  {\hat{\bf \phi}}~,~{\bf B}_g=\vec\nabla\times {\bf
v}_s={\hbar\over 2m_3} {{\bf r}\over r^3}  ~.
\label{Monopole1}
\end{equation}
On the $B$-phase side one has
\begin{equation}
{\bf B}_g=\vec\nabla\times {\bf v}_s={\pi\hbar\over m_3} \delta_2({\bf
\rho}) ~.
\label{Monopole2}
\end{equation}
The gravimagnetic flux propagates along the vortex in the B
phase towards the nexus (boojum) and then radially and
divergencelessly from the boojum into the A phase. This is the analog of the
gravimagnetic monopole discussed in
\cite{GravimagneticMonopole}.

\begin{figure}[!!!t]
\begin{center}
\leavevmode
\epsfig{file=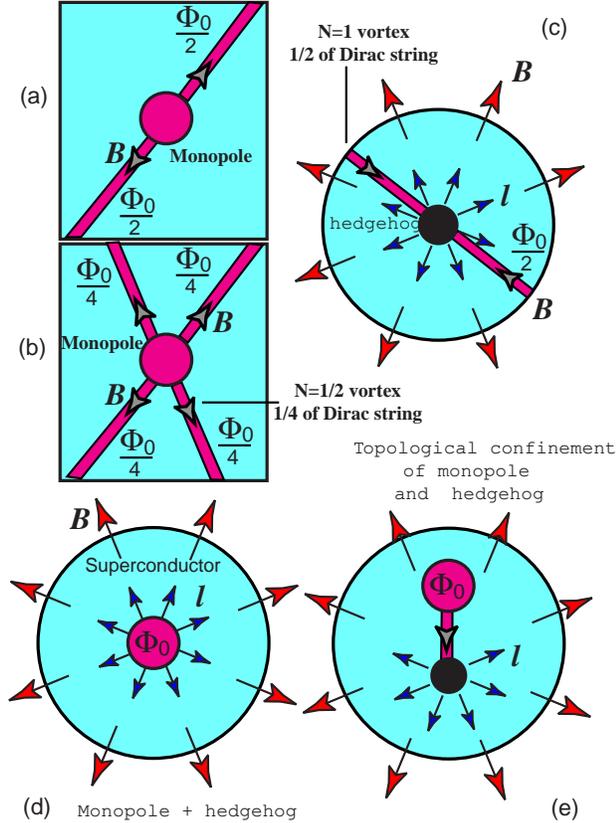,width=0.9\linewidth}
\caption[monopole]
    {Hedgehogs and magnetic monopoles in superconductors. (a)  `t
Hooft-Polyakov
magnetic monopole in conventional superconductor. Magnetic flux of the monopole
is concentrated in  Abrikosov vortices because of the Meissner effect; (b)
Magnetic monopole in a chiral superconductor with unifrom
$\hat{\bf l}$ vector. As distinct from the monopole inside the conventional
superconductors, the magnetic flux  $\Phi_0$ of the monopole can be carried
away
from the monopole by 4 half-quantum vortices.
(c) Nexus: hedgehog with two Abrikosov
vortices emanating from the core. Magnetic flux $\Phi_0$ enters the
core of the hedgehog
along two Abrikosov vortices with $N=1$ (or 4 vortices with $N=1/2$) and then
flows out radially along the lines of the $\hat{\bf
l}$-vector field.  (d)  `t Hooft-Polyakov  magnetic
monopole  + hedgehog in a chiral superconductor. The Abrikosov
vortices attached to the `t Hooft-Polyakov  magnetic monopole annihilate the
Abrikosov vortices attached to the hedgehog. Magnetic field
of the monopole penetrates radially into the bulk chiral superconductor
along the
lines of the $\hat{\bf l}$-vector field. (e) Topological confinement of the
`t Hooft-Polyakov  magnetic
monopole and hedgehog by Abrikosov strings in a chiral superconductor. For
simplicity a single Abrikosov string with $N=2$ is depicted. }
\label{MonopoleInChiral}
\end{center}
\end{figure}

\section{Topological interaction of magnetic monopoles with chiral
superconductors}

Since the `t Hooft-Polyakov magnetic monopole, which can exist in
GUT, and the monopole part of the nexus in chiral
superconductors have the same magnetic and topological charges, there is a
topological interaction between them. First, let us recall
what happens when the magnetic monopole enters a conventional superconductor:
because of the Meissner effect -- expulsion of the magnetic field from the
superconductor -- the magnetic field from the monopole will be concentrated in
two flux tubes of Abrikosov vortices with the total winding number
$N=2$ (Fig.~\ref{MonopoleInChiral}a).  In a chiral amorphous superconductor
these can form 4 flux tubes, represented by half-quantum Abrikosov vortices
(Fig.~\ref{MonopoleInChiral}b).

However,  the most interesting situation occurs if one takes into account that
in a chiral superconductor the Meissner effect is not complete because of the
$\hat{\bf l}$ texture. As we discussed above the
magnetic flux is not necessarily concentrated in the tubes, but can propagate
radially from the hedgehog (Fig.~\ref{MonopoleInChiral}c). If now the `t
Hooft-Polyakov magnetic monopole enters the core of the hedgehog in
Fig.~\ref{MonopoleInChiral}c, which has the same magnetic charge, their
strings,
i.e.   Abrikosov vortices carried by the monopole  and Abrikosov vortices
attached to the nexus, will annihilate each other. What is left is the combined
point defect: hedgehog + magnetic monopole without any attached strings
(Fig.~\ref{MonopoleInChiral}d). This means that the monopole destroys the
topological connection of the hedgehog and Abrikosov vortices, instead one has
 topological confinement between the monopole and hedgehog. The  core of the
hedgehog represents  the natural trap for one magnetic monopole:  if one
tries to
separate the monopole from the hedgehog, one must create the piece(s) of the
Abrikosov vortex(ices) which connect the hedgehog and the monopole
(Fig.~\ref{MonopoleInChiral}e).

\section{Discussion: Fermions in the presence of topological defects}

Fermions in  topologically nontrivial environments
behave in a curious way, especially in the presence of such exotic objects as
fractional vortices and monopoles discussed in this paper. In the presence of a
monopole the quantum statistics can change, for example, the isospin degrees of
freedom are trasformed to spin degrees\cite{JackiwRebbi}.  There are also
the so called fermion zero modes: the bound states at monopole or vortex,
which have exactly zero energy.  For example the $N=1/2$ vortex in a
two-dimensional chiral superconductor, which contains only one layer, has one
fermionic state with exactly zero energy
\cite{VolovikNexus}.  Since the zero-energy level
can be either filled or empty, there is a fractional  entropy $(1/2) \ln 2$ per
layer related to the vortex. The factor $(1/2)$ appears because in
superconductors the particle  excitation coincides with its antiparticle
(hole), {\it i.e.} the quasiparticle is a Majorana fermion (Nice discussion of
Majorana fermions in chiral superconductors can be found in
\cite{Read}).  Also the spin of the vortex in a chiral superconductor can be
fractional.  According to
\cite{Ivanov}, the $N=1$ vortex in a chiral superconductor must have a spin
$S=1/4$ (per layer); this implies the spin $S=1/8$ per layer for $N=1/2$
vortex.
Similarly there can be an anomalous fractional electric charge of the $N=1/2$
vortex, which is 1/2 of the fractional charge $e/4$ discussed for the $N=1$
vortex\cite{Goryo}. There is still some work to be done to elucidate the
problem with the fractional charge, spin and statistics, related to the
topological defects in chiral superconductors.

\vfill\eject

\end{document}